\documentclass[12pt,showpacs,english,showkeys]{article}
\usepackage{amsmath}
\usepackage{graphicx}
\usepackage{epsfig}
\usepackage{dcolumn}
\usepackage{bm}
\usepackage{slashed}
\usepackage{amsfonts}
\usepackage{amssymb}
\usepackage{fontenc}
\usepackage{amsmath}
\usepackage{fontenc}
\usepackage[english]{babel}
\usepackage{nameref}
\usepackage{epstopdf}

\input{tcilatex}
\begin{document}

\title{Unitarity and vacuum deformation in QED with critical potential steps}
\author{S.P. Gavrilov$^{1,2}$\thanks{%
gavrilovsergeyp@yahoo.com}, D. M. Gitman$^{1,3,4}$\thanks{%
gitman@if.usp.br} and A.A. Shishmarev $^{1,4}\thanks{%
a.a.shishmarev@mail.ru}$ \\
{\normalsize $^{1}$ Department of Physics, Tomsk State University, Tomsk
634050, Russia; }\\
{\normalsize $^{2}$ Department of General and Experimental Physics, }\\
{\normalsize Herzen State Pedagogical University of Russia,}\\
{\normalsize Moyka embankment 48, 191186 St. Petersburg, Russia;}\\
{\normalsize $^{3}$ P.N. Lebedev Physical Institute, 53 Leninsky prospekt,
119991 Moscow, Russia;}\\
{\normalsize $^{4}$ Institute of Physics, University of S\~{a}o Paulo, CP
66318, CEP 05315-970, S\~{a}o Paulo, SP, Brazil}\\
}
\maketitle

\begin{abstract}
The present article can be considered as a complement to the work P.R.D 
\textbf{93,} 045002 (2016) where an nonperturbative approach to QED with $x$%
-electric critical potential steps was developed. In the beginning we study
conditions when \textrm{in}- and \textrm{out}-spaces of the QED under
consideration are unitarily equivalent. Then we construct a general density
operator with the vacuum initial condition. Such an operator describes a
deformation of the initial vacuum state by $x$-electric critical potential
steps. We construct reductions of the deformed state to electron and
positron subsystems, calculating the loss of the information in these
reductions. We illustrate the general consideration studying the deformation
of the quantum vacuum between two capacitor plates. Finally we calculate the
entanglement measures of these reduced matrices as von Neumann entropies.
\end{abstract}

\section{Introduction}

Problems of quantum field theory with external field violating the vacuum
stability are already being studied systematically for a long time. Recently
they turned out to be of special attention due to new real possible
applications in astrophysics and physics of nanostructures. A
nonperturbative formulation of QED with the so-called $t$-potential electric
steps (time-depending potentials) was developed in Refs. \cite%
{Gitman1,Gitman2,Gitman3} and applied to various model and realistic
physical problems, see e.g. \cite{GGT,DvGavGi,GavGitSh15}. In the recent
work \cite{GavGi15} Gavrilov and Gitman succeeded to construct a consistent
version of QED with the so-called $x$-electric critical potential steps
(time-independent nonuniform electric fields of constant direction that are
concentrated in restricted space areas), for which a large area of new
important applications opens, see reviews in \cite{GavGi15,L-field}. However
many principle questions of the formulation still require detailed
clarification. The present work is devoted to some of them. In the beginning
we study conditions when \textrm{in}- and \textrm{out}-spaces of the QED
under consideration are unitarily equivalent. Then we construct a general
density operator with the vacuum initial condition. Such an operator
describes a deformation of the initial vacuum state by $x$-electric critical
potential steps. We construct reductions of the deformed state to electron
and positron subsystems, calculating the loss of the information in these
reductions. We illustrate the general consideration studying the deformation
of the quantum vacuum between two capacitor plates. In this article we
generally adapt the notations of the paper \cite{GavGi15}, where the general
theory of QED with $x$-electric critical potential steps was developed, and
Ref. \cite{L-field}, where the particular case of a constant electric field
between two capacitor plates was studied. In fact, the present article can
be considered as a complement to the work \cite{GavGi15}.

\section{Unitarity in QED with $x$-electric potential steps\label{Sec.2}}

It was shown in Ref. \cite{GavGi15} that in the presence of $x$-electric
potential steps the quantized Dirac field can be described in terms of 
\textrm{in} and \textrm{out} electrons and positrons. Such particles are
characterized by quantum numbers $n$ that can be divided in five ranges $%
\Omega _{i}$,\emph{\ }$i=1,...,5$. We denote\emph{\ }the corresponding
quantum numbers by $n_{i}$,\ so that $n_{i}\in \Omega _{i}$. The manifold of
all the quantum numbers $n$ is denoted by $\Omega $, so that $\Omega =\Omega
_{1}\cup \cdots \cup \ \Omega _{5}$. The \textrm{in-} and \textrm{out-}vacua
can be factorized%
\begin{equation}
\left\vert 0,\mathrm{in}\right\rangle =\sideset{}{^{\,\lower1mm\hbox{$%
\otimes$}}}\tprod\limits_{i=1}^{5}\left\vert 0,\mathrm{in}\right\rangle
^{\left( i\right) }\ ,\ \ \left\vert 0,\mathrm{out}\right\rangle =%
\sideset{}{^{\,\lower1mm\hbox{$\otimes$}}}\tprod\limits_{i=1}^{5}\left\vert
0,\mathrm{out}\right\rangle ^{\left( i\right) }\ ,  \label{2.1}
\end{equation}%
where $\left\vert 0,\mathrm{in}\right\rangle ^{\left( i\right) }$ and $%
\left\vert 0,\mathrm{out}\right\rangle ^{\left( i\right) }$ are the partial
vacua in the ranges $\Omega _{i}$. Note that in each range $\Omega _{i}$\ it
is also possible to factorize vacuum vectors in modes with fixed quantum
number $n$ so that%
\begin{equation}
\left\vert 0,\mathrm{in}\right\rangle ^{\left( i\right) }=\prod_{n\in \Omega
_{i}}\left\vert 0,\mathrm{in}\right\rangle _{n}^{\left( i\right) },\text{ \ }%
\left\vert 0,\mathrm{out}\right\rangle ^{\left( i\right) }=\prod_{n\in
\Omega _{i}}\left\vert 0,\mathrm{out}\right\rangle _{n}^{\left( i\right) }.
\label{2.1a}
\end{equation}

It was shown that all \textrm{in-} and \textrm{out-}vacua, except the vacua
in the range $\Omega _{3}$ (in the so-called Klein zone) coincide,%
\begin{equation}
\left\vert 0,\mathrm{out}\right\rangle ^{\left( i\right) }=\left\vert 0,%
\mathrm{in}\right\rangle ^{\left( i\right) },\ \ i=1,2,4,5,\ \ \left\vert 0,%
\mathrm{out}\right\rangle ^{\left( 3\right) }\neq \left\vert 0,\mathrm{in}%
\right\rangle ^{\left( 3\right) }.  \label{2.2}
\end{equation}

In what follows, we use the subindex $K$ to denote all the quantities from
the Klein zone, e.g. $\left\vert 0,\mathrm{in}\right\rangle ^{\left(
3\right) }=\left\vert 0,\mathrm{in}\right\rangle ^{\left( K\right) }$, $%
\Omega _{3}=\Omega _{K},$ and so on.

The vacuum-to-vacuum transition amplitude $c_{v}=\langle 0,\mathrm{out}|0,%
\mathrm{in}\rangle $ coincides (due to Eq. (\ref{2.2})) with the
vacuum-to-vacuum transition amplitude $c_{v}^{\left( K\right) }$ in the
Klein zone,%
\begin{equation}
c_{v}=\langle 0,\mathrm{out}|0,\mathrm{in}\rangle =c_{v}^{\left( K\right)
}=\ ^{\left( K\right) }\langle 0,\mathrm{out}|0,\mathrm{in}\rangle ^{(K)}\ .
\label{2.4}
\end{equation}

The linear canonical transformation between the \textrm{in} and \textrm{out}
sets of creation and annihilation operators in the Klein zone ($a$ and $b$
operators are related to electrons and positrons, respectively) can be
written in the following form%
\begin{align}
\ & ^{-}a_{n}(\mathrm{in})=w_{n}\left( +|+\right) ^{-1}\left[ \text{ }%
^{+}a_{n}(\mathrm{out})+w_{n}\left( +-|0\right) \ _{+}b_{n}^{\dagger }(%
\mathrm{out})\right] ,  \notag \\
& \ _{-}b_{n}^{\dagger }(\mathrm{in})=w_{n}\left( -|-\right) ^{-1}\left[ \
_{+}b_{n}^{\dagger }(\mathrm{out})-w_{n}\left( +-|0\right) \text{ }^{+}a_{n}(%
\mathrm{out})\right] ,  \label{2.3a}
\end{align}%
where 
\begin{eqnarray}
&&w\left( +|+\right) _{n^{\prime }n}=c_{v}^{-1}\langle 0,\mathrm{out}%
\left\vert \ ^{+}a_{n^{\prime }}\left( \mathrm{out}\right) \
^{-}a_{n}^{\dagger }(\mathrm{in})\right\vert 0,\mathrm{in}\rangle ,  \notag
\\
&&w\left( -|-\right) _{n^{\prime }n}=c_{v}^{-1}\langle 0,\mathrm{out}%
\left\vert \ _{+}b_{n^{\prime }}\left( \mathrm{out}\right) \
_{-}b_{n}^{\dagger }(\mathrm{in})\right\vert 0,\mathrm{in}\rangle \,,
\label{2.5}
\end{eqnarray}%
are relative scattering amplitudes of electrons and positrons, and%
\begin{eqnarray}
&&w\left( +-|0\right) _{n^{\prime }n}=c_{v}^{-1}\langle 0,\mathrm{out}%
\left\vert \ ^{+}a_{n^{\prime }}\left( \mathrm{out}\right) \ _{+}b_{n}\left( 
\mathrm{out}\right) \right\vert 0,\mathrm{in}\rangle \,,  \notag \\
&&w\left( 0|-+\right) _{nn^{\prime }}=c_{v}^{-1}\langle 0,\mathrm{out}%
\left\vert \ _{-}b_{n}^{\dagger }(\mathrm{in})\ ^{-}a_{n^{\prime }}^{\dagger
}(\mathrm{in})\right\vert 0,\mathrm{in}\rangle \,.  \label{2.6}
\end{eqnarray}%
are relative amplitudes of a pair creation and a pair annihilation, and%
\begin{equation}
c_{v}=c_{v}^{\left( K\right) }=\dprod\limits_{n}w_{n}\left( -|-\right)
^{-1}\,.  \label{2.7}
\end{equation}%
All the amplitude can be expressed via the coefficients $g\left( _{\zeta
}\left\vert ^{\zeta ^{\prime }\ }\right. \right) $ which, in turn, are
calculated via corresponding solutions of the Dirac equation with $x$%
-electric potential steps.

An important question is whether \textrm{in} and \textrm{out} spaces are
unitarily equivalent? The answer is positive if the linear canonical
transformation (\ref{2.3a}) (together with its adjoint transformation) is
proper one. In the latter case there exists a unitary operator $V$, such that%
\begin{eqnarray}
&&V\left( a(\mathrm{out}),a^{\dag }(\mathrm{out}),b(\mathrm{out}),b^{\dag }(%
\mathrm{out})\right) V^{\dag }=\left( a(\mathrm{in}),a^{\dag }(\mathrm{in}%
),b(\mathrm{in}),b^{\dag }(\mathrm{in})\right) ,  \notag \\
&&\left\vert 0,\mathrm{in}\right\rangle =V\left\vert 0,\mathrm{out}%
\right\rangle ,\ V^{\dag }=V^{-1}\ .  \label{2.8}
\end{eqnarray}

Let us denote all the \textrm{out} operators via $\alpha $ and all the 
\textrm{in} operators via $\beta .$ Then the linear uniform canonical
transformation between these operators can be written as (we consider the
only Fermi case here)%
\begin{equation}
\beta =\Phi \alpha +\Psi \alpha ^{+},\ \ \Phi \Phi ^{+}+\Psi \Psi ^{+}=1,\
\Phi \Psi ^{T}+\Psi \Phi ^{T}=0.  \label{2.9}
\end{equation}%
According to (\cite{Berezin,Kiperman}), transformation (\ref{2.9}) is proper
one if $\Psi $ is a Hilbert-Schmidt operator, i.e., $\dsum\limits_{m,n}\left%
\vert \Psi _{mn}\right\vert ^{2}<\infty $. It is easily to see that
Hilbert-Schmidt criterion for the transformation (\ref{2.3a}) reads%
\begin{equation}
\sum\limits_{n}\left[ \left\vert \frac{w_{n}\left( +-|0\right) }{w_{n}\left(
+|+\right) }\right\vert ^{2}+\left\vert \frac{w_{n}\left( +-|0\right) }{%
w_{n}\left( -|-\right) }\right\vert ^{2}\right] <\infty .  \label{2.10}
\end{equation}

As it was shown in Ref. \cite{GavGi15},%
\begin{equation}
\left\vert \frac{w_{n}\left( +-|0\right) }{w_{n}\left( +|+\right) }%
\right\vert ^{2}=N_{n}^{a},\ \left\vert \frac{w_{n}\left( +-|0\right) }{%
w_{n}\left( -|-\right) }\right\vert ^{2}=N_{n}^{b},  \label{2.11}
\end{equation}%
where $N_{n}^{a}$ and $N_{n}^{b}$ are differential mean numbers of electrons
and positrons created from the vacuum by the potential step. Then, the
left-hand side of Eq.~(\ref{2.10}) is the total number $N$ of particles
created from the vacuum, such that unitarity condition can be written as%
\begin{equation}
\sum\limits_{n}\left( N_{n}^{a}+N_{n}^{b}\right) =N<\infty .  \label{2.12}
\end{equation}%
Note that in- and out-spaces of the scalar QED in the presence of critical
potential steps are unitarily equivalent under the same condition. 

For realistic external field limited in space and time this condition is
obviously satisfied.

Inequality (\ref{2.10}) derived for QED with $x$-electric potential steps
can be considered as one more confirmation of the consistency of the latter
theory and correct interpretation of \textrm{in} and \textrm{out} particles
there. One should note that qualitatively similar result was established in
Ref. \cite{Gitman2} for QED with time-dependent electric potential steps.

\section{Deformation of initial vacuum state\label{Sec.3}}

In this section we are going to study deformation of initial vacuum state
under the action of a $x$-electric potential step.

In the Heisenberg picture, the density operator of the system whose initial
state is the vacuum, is given by equation%
\begin{equation}
\hat{\rho}=|0,\mathrm{in}\rangle \langle 0,\mathrm{in}|.  \label{3.1}
\end{equation}%
The \textrm{in} and \textrm{out} Fock spaces are related by the unitary
operator $V$,\ see (\ref{2.8}). Then%
\begin{equation}
\hat{\rho}=V|0,\mathrm{out}\rangle \langle 0,\mathrm{out}|V^{\dag }\ .
\label{3.2}
\end{equation}

In QED with $x$-electric potential steps the operator $V$ was constructed in 
\cite{GavGi15}. Since it can be factorized, the density operator (\ref{3.2})
can be factorized as well, 
\begin{eqnarray}
&&V=\prod\limits_{i=1}^{5}V^{(i)},\ \ |0,\mathrm{in}\rangle ^{(i)}=V^{(i)}|0,%
\mathrm{out}\rangle ^{(i)},  \notag \\
&&\hat{\rho}=\prod\limits_{i=1}^{5}V^{(i)}|0,\mathrm{out}\rangle ^{\left(
i\right) }\;^{(i)}\langle 0,\mathrm{out}|V^{(i)\dag }.  \label{3.3}
\end{eqnarray}%
Due to the specific structure of the operator $V^{\left( i\right) }$,$\
i=1,2,4,5$,\ we have%
\begin{equation*}
V^{(i)}|0,\mathrm{out}\rangle ^{\left( i\right) }\;^{\left( i\right)
}\langle 0,\mathrm{out}|V^{(i)\dag }=|0,\mathrm{out}\rangle ^{\left(
i\right) }\;^{(i)}\langle 0,\mathrm{out}|\ =|0,\mathrm{in}\rangle ^{\left(
i\right) }\;^{(i)}\langle 0,\mathrm{in}|,\ \ i=1,2,4,5.
\end{equation*}%
The latter relation has clear physical meaning\ --\ vacuum states in the
ranges $\Omega _{1}$, $\Omega _{2}$, $\Omega _{4}$, and $\Omega _{5}$ do not
change with time, there is no particle creation there. Let us use the
following notation%
\begin{eqnarray}
P^{\prime } &=&\prod\limits_{i=1,2,4,5}|0,\mathrm{out}\rangle ^{\left(
i\right) }\;^{(i)}\langle 0,\mathrm{out}|\ =\prod\limits_{i=1,2,4,5}|0,%
\mathrm{in}\rangle ^{\left( i\right) }\;^{(i)}\langle 0,\mathrm{in}|,  \notag
\\
\ \hat{\rho}_{K} &=&V^{(K)}P_{K}V^{(K)\dag },\ \ P_{K}=|0,\mathrm{out}%
\rangle ^{\left( K\right) }\;^{(K)}\langle 0,\mathrm{out}|,  \label{3.4}
\end{eqnarray}%
then%
\begin{equation}
\hat{\rho}=P^{\prime }\hat{\rho}_{K}\ .  \label{3.5}
\end{equation}

Using the following explicit form of the operator $V^{(K)}=V^{(3)}$ derived
in Ref. \cite{GavGi15},%
\begin{align*}
V^{(K)}& =\exp \left[ -\sum_{n\in \Omega _{K}}{\ }^{+}a_{n}^{\dag }(\mathrm{%
out})w_{n}\left( +-|0\right) {\ }_{+}b_{n}^{\dag }(\mathrm{out})\right] \\
& \times \exp \left[ -\sum_{n\in \Omega _{K}}{\ }_{+}b_{n}(\mathrm{out})\ln
w_{n}\left( -|-\right) {\ }_{+}b_{n}^{\dag }(\mathrm{out})\right] \\
& \times \exp \left[ \sum_{n\in \Omega _{K}}{\ }^{+}a_{n}^{\dag }(\mathrm{out%
})\ln w_{n}\left( +|+\right) {\ }^{+}a_{n}(\mathrm{out})\right] \\
& \times \exp \left[ -\sum_{n\in \Omega _{K}}{\ }_{+}b_{n}(\mathrm{out}%
)w_{n}\left( 0|-+\right) {\ }^{+}a_{n}(\mathrm{out})\right] ,
\end{align*}%
one can derive two alternative expressions for the density operator $\hat{%
\rho}_{K}$.

The first one is a normal form exponential with respect to the \textrm{out}%
-operators (denoted by $:\ldots :$): 
\begin{gather}
\hat{\rho}_{K}|c_{v}|^{-2}=\mathbf{:}\exp \left\{ -\sum_{n\in \Omega _{K}}%
\left[ \text{ }^{+}a_{n}^{\dag }(\mathrm{out})\text{ }^{+}a_{n}(\mathrm{out}%
)+\text{ }_{+}b_{n}^{\dag }(\mathrm{out})\text{\ }_{+}b_{n}(\mathrm{out}%
)\right. \right.  \notag \\
+\left. \left. \text{ }^{+}a_{n}^{\dag }(\mathrm{out})w_{n}\left(
+-|0\right) \text{ }_{+}b_{n}^{\dag }(\mathrm{out})+\text{ }_{+}b_{n}(%
\mathrm{out})w_{n}\left( +-|0\right) ^{\ast }\text{ }^{+}a_{n}(\mathrm{out})%
\right] \right\} \mathbf{:\ }.  \label{3.7}
\end{gather}%
Representation (\ref{3.7}) can be derived in the following way: Using (\ref%
{3.4}) and the explicit form of $V^{(K)},$ we can write\emph{\ }%
\begin{eqnarray}
&&\hat{\rho}_{K}|c_{v}|^{-2}=\exp \left[ -\sum_{n\in \Omega _{K}}{\ }%
^{+}a_{n}^{\dag }(\mathrm{out})w_{n}\left( +-|0\right) {\ }_{+}b_{n}^{\dag }(%
\mathrm{out})\right]  \notag \\
&&P_{K}\exp \left[ -\sum_{n\in \Omega _{K}}{\ }_{+}b_{n}(\mathrm{out}%
)w_{n}\left( +-|0\right) ^{\ast }{\ }^{+}a_{n}(\mathrm{out})\right] .
\label{3.7c}
\end{eqnarray}%
Making use of well-known Berezin representation \cite{Berezin} for a
projection operator\emph{\ }$P_{K}$ on the vacuum state, 
\begin{equation}
\ P_{K}=\mathbf{:}\exp \left\{ -\sum_{n\in \Omega _{K}}\left[ \text{ }%
^{+}a_{n}^{\dag }(\mathrm{out})\text{ }^{+}a_{n}(\mathrm{out})+\text{ }%
_{+}b_{n}^{\dag }(\mathrm{out})\text{\ }_{+}b_{n}(\mathrm{out})\right]
\right\} \mathbf{:}  \label{3.7b}
\end{equation}%
and taking into account that the left and the right exponents in Eq. (\ref%
{3.7c}) are already normal ordered, we easily obtain representation (\ref%
{3.7}).

The second representation reads:%
\begin{eqnarray}
&&\ \hat{\rho}_{K}|c_{v}|^{-2}=\prod_{n\in \Omega _{K}}\left[ 1-{\ }%
^{+}a_{n}^{\dag }(\mathrm{out})w_{n}\left( +-|0\right) {\ }_{+}b_{n}^{\dag }(%
\mathrm{out})\right]  \notag \\
&&\times P_{K,n}\left[ 1-{\ }_{+}b_{n}(\mathrm{out})w_{n}\left( +-|0\right)
^{\ast }{\ }^{+}a_{n}(\mathrm{out})\right] ,\text{ \ }  \notag \\
\text{ } &&P_{K,n}=|0,\mathrm{out}\rangle _{n}^{(K)}\ {}_{n}^{(K)}\langle 0,%
\mathrm{out}|.  \label{3.8a}
\end{eqnarray}%
Representation (\ref{3.8a})\ can be derived as follows: Using the fact that
operators with different quantum numbers $n$\ commute, and using the
relation, see, e.g., Ref. \cite{GGT}\emph{,\ }%
\begin{equation}
\exp \left[ a^{\dag }Da\right] =\mathbf{:}\exp \left[ a^{\dag }\left(
e^{D}-1\right) a\right] \mathbf{:\ },  \label{3.9}
\end{equation}%
to transform exponents from $V^{(K)}$,\ we expand then the obtained
expressions in power series. Since the \textrm{out}-operators in $V^{(K)}$
are Fermi type, these series are reduced to finite term expressions. Their
actions on the vacuum $|0,\mathrm{out}\rangle ^{(K)}$\ can be easily
calculated, and using of Eq.~(\ref{2.1a}), we arrive at Eq.~(\ref{3.8a}).

Finally we consider the structure of the $|0,\mathrm{in}\rangle $ state in
terms of \textrm{out}-operators. First of all we use the fact that the state
vector under discussion is factorized,%
\begin{eqnarray}
&&|0,\mathrm{in}\rangle =V|0,\mathrm{out}\rangle =|0,\mathrm{in}\rangle
^{\prime }|0,\mathrm{in}\rangle ^{(K)},  \notag \\
&&|0,\mathrm{in}\rangle ^{\prime }=\prod\limits_{i=1,2,4,5}|0,\mathrm{in}%
\rangle ^{(i)},\ \ |0,\mathrm{in}\rangle ^{(K)}=V^{(K)}|0,\mathrm{out}%
\rangle ^{(K)}.  \label{3.10}
\end{eqnarray}%
Then using the explicit form $V^{(K)}$, we obtain%
\begin{equation}
|0,\mathrm{in}\rangle ^{(K)}=c_{v}\prod\limits_{n\in \Omega _{K}}\left[ 1-{\ 
}^{+}a_{n}^{\dag }(\mathrm{out})w_{n}\left( +-|0\right) {\ }_{+}b_{n}^{\dag
}(\mathrm{out})\right] |0,\mathrm{out}\rangle ^{(K)}.  \label{3.11}
\end{equation}

In each fixed mode $n\in \Omega _{K}$, the state vector $|0,\mathrm{in}%
\rangle $ is a linear superposition of two terms -- the vacuum vector in
this mode and a state with an electron-positron pair.\emph{\ }

\section{Reductions to electron and positron subsystems\label{Sec.4}}

It should be stressed that the system under consideration can be considered
as a composed from a subsystem of electrons and a subsystem of positrons.
One can introduce the so-called two reduced density operators: $\hat{\rho}%
_{+}$ of the electron subsystem and $\hat{\rho}_{-}$ of the positron
subsystem, averaging complete density operator (\ref{3.1}) over all possible
positron states or over all possible electron states, respectively,

\begin{align}
& \hat{\rho}_{+}=\mathrm{tr}_{-}\hat{\rho}=\sum_{i=3}^{5}\sum_{M}\sum_{\{m\}%
\in \Omega _{i}}{}_{b}^{(i)}\langle M,\mathrm{out}|\hat{\rho}|M,\mathrm{out}%
\rangle _{b}^{(i)}\,,  \notag \\
& \hat{\rho}_{-}=\mathrm{tr}_{+}\hat{\rho}=\sum_{i=1}^{3}\sum_{M}\sum_{\{m\}%
\in \Omega _{i}}{}_{a}^{(i)}\langle M,\mathrm{out}|\hat{\rho}|M,\mathrm{out}%
\rangle _{a}^{(i)}\,,  \notag \\
& |M,\mathrm{out}\rangle _{b}^{(i)}=\left( M!\right)
^{-1/2}b_{m_{1}}^{\dagger }(\mathrm{out})\ldots b_{m_{M}}^{\dagger }(\mathrm{%
out})|0,\mathrm{out}\rangle _{b}^{(i)},  \notag \\
& |M,\mathrm{out}\rangle _{a}^{(i)}=\left( M!\right)
^{-1/2}a_{m_{1}}^{\dagger }(\mathrm{out})\ldots a_{m_{M}}^{\dagger }(\mathrm{%
out})|0,\mathrm{out}\rangle _{a}^{(i)}.  \label{reduction}
\end{align}%
Vectors $|0,\mathrm{out}\rangle _{a}^{(i)}$ and $|0,\mathrm{out}\rangle
_{b}^{(i)}$ are the electron and positron vacua in $\Omega _{i}$-range,
defined by 
\begin{equation}
{a}_{n}^{(i)}(\mathrm{out})|0,\mathrm{out}\rangle _{a}^{(i)}=0,\text{ }\ {b}%
_{n}^{(i)}(\mathrm{out})|0,\mathrm{out}\rangle _{b}^{(i)}=0,  \label{4.1}
\end{equation}%
where ${a}_{n}^{(i)}(\mathrm{out})$ and ${b}_{n}^{(i)}(\mathrm{out})$ are
corresponding annihilation operators of electron and positron in this range,
respectively. Of course, these electron and positron vacua can be factorized
in quantum modes, as was mentioned already above. One can see that%
\begin{eqnarray}
&|0,\mathrm{out}\rangle ^{(1,2)}=&|0,\mathrm{out}\rangle
_{a}^{(1,2)}=\prod_{n\in \Omega _{1,2}}|0,\mathrm{out}\rangle _{n,a}^{(1,2)},
\notag \\
&|0,\mathrm{out}\rangle ^{(4,5)}=&|0,\mathrm{out}\rangle
_{b}^{(4,5)}=\prod_{n\in \Omega _{4,5}}|0,\mathrm{out}\rangle _{n,b}^{(4,5)},
\notag \\
&|0,\mathrm{out}\rangle ^{(3)}=&|0,\mathrm{out}\rangle ^{(K)}=|0,\mathrm{out}%
\rangle _{a}^{(K)}\otimes |0,\mathrm{out}\rangle _{b}^{(K)},  \notag \\
&|0,\mathrm{out}\rangle _{a}^{(K)}=&\prod_{n\in \Omega _{K}}|0,\mathrm{out}%
\rangle _{n,a}^{(K)},\text{ \ }|0,\mathrm{out}\rangle _{b}^{(K)}=\prod_{n\in
\Omega _{K}}|0,\mathrm{out}\rangle _{n,b}^{(K)}.  \label{4.1a}
\end{eqnarray}

Using Eq. (\ref{3.5}) and representation (\ref{3.8a}) for $\hat{\rho}_{K}$,
it is easy to calculate traces in Eqs. (\ref{reduction}), and to obtain thus
explicit forms of the reduced operators $\hat{\rho}_{\pm }$:%
\begin{eqnarray}
&&\hat{\rho}_{+}|c_{v}|^{-2}=\prod\limits_{i=1,2}|0,\mathrm{out}\rangle
^{(i)}\ {}^{(i)}\langle 0,\mathrm{out}|  \notag \\
&&\otimes \prod_{n\in \Omega _{K}}\left[ P_{K,a,n}+|w_{n}\left( +-|0\right)
|^{2}{}^{+}a_{n}^{\dag }(\mathrm{out})P_{K,a,n}{}^{+}a_{n}(\mathrm{out})%
\right] ,  \notag \\
&&\hat{\rho}_{-}|c_{v}|^{-2}=\prod\limits_{i=4,5}|0,\mathrm{out}\rangle
^{(i)}\ {}^{(i)}\langle 0,\mathrm{out}|  \notag \\
&&\otimes \prod_{n\in \Omega _{K}}\left[ P_{K,b,n}+|w_{n}\left( +-|0\right)
|^{2}\text{ }_{+}b_{n}^{\dag }(\mathrm{out})P_{K,b,n}\text{{}}_{+}b_{n}(%
\mathrm{out})\right] ,  \notag \\
&&P_{K,a,n}=|0,\mathrm{out}\rangle _{n,a}^{(K)}\ {}_{n,a}^{(K)}\langle 0,%
\mathrm{out}|,\text{ \ \ }P_{K,b,n}=|0,\mathrm{out}\rangle _{n,b}^{(K)}\
{}_{n,b}^{(K)}\langle 0,\mathrm{out}|.  \label{4.2}
\end{eqnarray}

We can also consider a reduction of density operator (\ref{3.5}), which
occur due to measurement of a physical quantity by some classical tool, or,
in other words, due to decoherence. Suppose that we are measuring the number
of particles $N(\mathrm{out})$ in the state $\hat{\rho}$ of the system under
consideration. The operator corresponding to this physical quantity is $\hat{%
N}(\mathrm{out})=\sum_{i=1}^{5}\hat{N}_{i}(\mathrm{out}),$ where%
\begin{eqnarray}
&&\hat{N}_{1}(\mathrm{out})=\sum_{n\in \Omega _{1}}\left[ \text{ }%
^{+}a_{n}^{\dag }(\mathrm{out})\text{ }^{+}a_{n}(\mathrm{out})+\text{ }%
_{-}a_{n}^{\dag }(\mathrm{out})\text{ }_{-}a_{n}(\mathrm{out})\right] , 
\notag \\
&&\hat{N}_{2}(\mathrm{out})=\sum_{n\in \Omega _{2}}a_{n}^{\dag }a_{n},\text{
\ }\hat{N}_{4}(\mathrm{out})=\sum_{n\in \Omega _{4}}b_{n}^{\dag }b_{n}, 
\notag \\
&&\hat{N}_{3}(\mathrm{out})=\sum_{n\in \Omega _{K}}\left[ {\ }%
^{+}a_{n}^{\dag }(\mathrm{out}){\ }^{+}a_{n}^{\dag }(\mathrm{out})+{\ }%
_{+}b_{n}^{\dag }(\mathrm{out}){\ }_{+}b_{n}(\mathrm{out})\right] ,  \notag
\\
&&\hat{N}_{5}(\mathrm{out})=\sum_{n\in \Omega _{5}}\left[ \text{ }%
_{+}b_{n}^{\dag }(\mathrm{out})\text{ }_{+}b_{n}(\mathrm{out})+\text{ }%
^{-}b_{n}^{\dag }(\mathrm{out})\text{ }^{-}b_{n}(\mathrm{out})\right] .
\label{4.3}
\end{eqnarray}

According to von Neumann \cite{Neumann}, the density operator $\hat{\rho}$
after such a measurement is reduced to the operator $\hat{\rho}_{N}$ of a
form 
\begin{equation}
\hat{\rho}_{N}=\sum_{s}\langle s,\mathrm{out}|\hat{\rho}|s,\mathrm{out}%
\rangle \hat{P}_{s},\text{ \ }\hat{P}_{s}=|s,\mathrm{out}\rangle \langle s,%
\mathrm{out}|,  \label{4.4}
\end{equation}%
where $|s,\mathrm{out}\rangle $ are eigenstates of the operator $\hat{N}(%
\mathrm{out})$ with the eigenvalues $s$ that represent the total number of
electrons and positrons in the state $|s,\mathrm{out}\rangle $, 
\begin{eqnarray*}
&&\hat{N}(\mathrm{out})|s,\mathrm{out}\rangle =s|s,\mathrm{out}\rangle , \\
&&\ |s,\mathrm{out}\rangle =\prod_{n\in \Omega _{1}}\left[ \text{ }%
^{+}a_{n}^{\dag }(\mathrm{out})\right] ^{l_{n,1}}\left[ \text{ }%
_{-}a_{n}^{\dag }(\mathrm{out})\right] ^{k_{n,1}}\prod_{n\in \Omega
_{2}}\left( \text{ }a_{n}^{\dag }\right) ^{l_{n,2}}\prod_{n\in \Omega
_{4}}\left( \text{ }b_{n}^{\dag }\right) ^{l_{n,4}} \\
&&\times \prod_{n\in \Omega _{5}}\left[ \text{ }_{+}b_{n}^{\dag }(\mathrm{out%
})\right] ^{l_{n,5}}\left[ \text{ }^{-}b_{n}^{\dag }(\mathrm{out})\right]
^{k_{n,5}}\prod_{n\in \Omega _{K}}\left[ {\ }^{+}a_{n}^{\dag }(\mathrm{out})%
\right] ^{l_{n,3}}\left[ {\ }_{+}b_{n}^{\dag }(\mathrm{out})\right]
^{k_{n,3}}|0,\mathrm{out}\rangle , \\
&&s=\sum_{n\in \Omega _{1}}\left( l_{n,1}+k_{n,1}\right) +\sum_{n\in \Omega
_{2}}\left( l_{n,2}\right) +\sum_{n\in \Omega _{4}}\left( l_{n,4}\right)
+\sum_{n\in \Omega _{5}}\left( l_{n,5}+k_{n,5}\right) +\sum_{n\in \Omega
_{K}}\left( l_{n,3}+k_{n,3}\right) .
\end{eqnarray*}%
Note that $l_{n,i}$, $k_{n,i}=(0,1),$ due to the fact that we deal with
fermions.

Due to the structure of the operator $\hat{\rho}$, the weights $\langle s,%
\mathrm{out}|\hat{\rho}|s,\mathrm{out}\rangle $ are nonzero only for pure
states $|s,\mathrm{out}\rangle $ with an integer number of pairs in $\Omega
_{K}$ (since the initial state of the system was a vacuum, and there is no
particle creation outside of the Klein zone). Thus, the operator $\hat{\rho}%
_{N}$ takes the form%
\begin{equation}
\hat{\rho}_{N}|c_{v}|^{-2}=P^{\prime }\prod_{n\in \Omega _{K}}\left[
P_{K,n}+\!|w_{n}\left( +-|0\right) |^{2}\text{ }^{+}a_{n}^{\dag }(\mathrm{out%
})\text{ }_{+}b_{n}^{\dag }(\mathrm{out})P_{K,n}\text{ }_{+}b_{n}(\mathrm{out%
})\text{ }^{+}a_{n}(\mathrm{out})\right] ,  \label{4.8}
\end{equation}%
where operators $P_{K,n}$ and $P^{\prime }$ were defined in the previous
Section, see Eq. (\ref{3.8a}). Note that the measurement destroys
nondiagonal terms of the density operator (\ref{3.8a}).

Since the operator $V$ is unitary and the initial state of the system under
consideration is a pure state (the vacuum state) the density operator (\ref%
{3.5}) describes a pure state as well. Therefore its von Neumann entropy is
zero. However, the reduced density operators $\hat{\rho}_{\pm }$ (\ref{4.2})
describe already mixed states and their entropies $S(\hat{\rho}_{\pm })$ are
not zero,%
\begin{equation}
S(\hat{\rho}_{\pm })=-k_{B}\mathrm{tr}\hat{\rho}_{\pm }\ln \hat{\rho}_{\pm }.
\label{4.9}
\end{equation}%
It is known that this entropy can be treated as a measure of the quantum
entanglement of the electron and positron subsystems and can be treated as
the measure of the information loss.

Using the normalization condition for the reduced density operators, $%
\mathrm{tr}\hat{\rho}_{\pm }=1$, the relation (\ref{3.9}), definitions for
differential mean numbers of particles $N_{n}^{a}$\ and antiparticles $%
N_{n}^{b}$\ created from vacuum 
\begin{equation}
N_{n}^{a}=\mathrm{tr}\hat{\rho}_{+}a_{n}^{\dagger }(\mathrm{out})a_{n}(%
\mathrm{out}),\ N_{n}^{b}=\mathrm{tr}\hat{\rho}_{-}b_{n}^{\dagger }(\mathrm{%
out})b_{n}(\mathrm{out}),  \label{4.11}
\end{equation}%
and the fact that%
\begin{equation}
N_{n}^{a}=N_{n}^{b}=N_{n}^{\text{$\mathrm{cr}$}},\ \ |w_{n}\left(
+-|0\right) |^{2}=N_{n}^{\text{$\mathrm{cr}$}}\left( 1-N_{n}^{\text{$\mathrm{%
cr}$}}\right) ^{-1},  \label{4.12}
\end{equation}%
we can calculate traces in Eqs. (\ref{4.9}) and rewrite RHS in these
equations as 
\begin{equation}
S(\hat{\rho}_{\pm })=\sum_{n\in \Omega _{K}}S_{n},\text{ \ }S_{n}=-k_{B}%
\left[ (1-N_{n}^{\text{$\mathrm{cr}$}})\ln \left( 1-N_{n}^{\text{$\mathrm{cr}
$}}\right) +N_{n}^{\text{$\mathrm{cr}$}}\ln N_{n}^{\text{$\mathrm{cr}$}}%
\right] .  \label{4.13}
\end{equation}

The von Neumann-reduced density operator (\ref{4.8}) also describe mixed
state; making use of the fact that the pure states $|0,\mathrm{out}\rangle
_{n}^{(K)}$ and $\text{ }^{+}a_{n}^{\dag }(\mathrm{out})\text{ }%
_{+}b_{n}^{\dag }(\mathrm{out})|0,\mathrm{out}\rangle _{n}^{(K)}$ are
orthogonal and normalized, it is not difficult to show that the von Neumann
entropy $S(\hat{\rho}_{N})$ of the mixed state (\ref{4.8}) coincide with the
entropies $S(\hat{\rho}_{\pm })$ of the reduced density operators $\hat{\rho}%
_{\pm }$.

The differential mean number of fermions created $N_{n}^{\text{$\mathrm{cr}$}%
}$ can vary only within the range $(0,1)$. The partial entropy $S_{n}$ for
given $n$ in Eq.~(\ref{4.13}) is symmetric with respect to value of $N_{n}^{%
\text{$\mathrm{cr}$}}$. It reaches maximum at $N_{n}^{\text{$\mathrm{cr}$}%
}=1/2$ and turns to zero at $N_{n}^{\text{$\mathrm{cr}$}}=1$ and $N_{n}^{%
\text{$\mathrm{cr}$}}=0$. This fact can be interpreted as follows. In the
case of $N_{n}^{\text{$\mathrm{cr}$}}=0$ there are no particles created by
the external field and the initial vacuum state in the mode remains
unchanged. The case $N_{n}^{\text{$\mathrm{cr}$}}=1$ corresponds to the
situation when a particle is created with certainty. The maximum of  $S_{n}$%
, corresponding to $N_{n}^{\text{$\mathrm{cr}$}}=1/2$, is associated with
the state with the maximum amount of uncertainty.

\section{Deformation of the quantum vacuum between two capacitor plates 
\label{Sec.6}}

Here we illustrate the general consideration considering the deformation of
the quantum vacuum between two infinite capacitor plates separated by a
finite distance $L$. Some aspects of particle creation by the constant
electric field between such plates (this field is also called $L$-constant
electric field) were studied in Ref. \cite{L-field}. The latter field is a
particular case of $x$-electric potential step. Thus, we consider the $L$%
-constant electric field in $d=D+1$ dimensions. We chose $\mathbf{E}%
(x)=\left( E^{i},\ i=1,...,D\right) ,\ E^{1}=E_{x}(x),\ E^{2,...,D}=0$,%
\begin{equation*}
E_{x}(x)=\left\{ 
\begin{array}{l}
0,\ x\in (-\infty ,-L/2] \\ 
E=\mathrm{const}>0,\ x\in (-L/2,L/2) \\ 
0,\ x\in \lbrack L/2,\infty )%
\end{array}%
\right. .
\end{equation*}%
The potential energy of an electron in the $L$-electric field under
consideration is 
\begin{equation}
U(x)=\left\{ 
\begin{array}{ll}
U_{\mathrm{L}}=-eEL/2, & x\in (-\infty ,-L/2] \\ 
eEx, & x\in (-L/2,L/2) \\ 
U_{\mathrm{R}}=eEL/2, & x\in \lbrack L/2,\infty )%
\end{array}%
\right. .  \label{6.2}
\end{equation}%
The magnitude of the corresponding $x$-electric is $\mathbb{U}=eEL.$ We are
interested in the critical steps, for which 
\begin{equation}
\mathbb{U}=eEL>2m  \label{6.2a}
\end{equation}%
and the vacuum is unstable in the Klein zone.

We consider a particular case with a sufficiently large length $L$ between
the capacitor plates, {\large \ }%
\begin{equation}
\sqrt{eE}L\gg \max \left\{ 1,E_{c}/E\right\} .  \label{L-large}
\end{equation}%
Here $E_{c}=m^{2}/e$\ is the critical Schwinger field. In what follows we
conditionally call this approximation as large work approximation. Such%
{\large \ }kind of $x$-electric step represent a regularization for a
constant uniform electric field and is suitable for imitating a
small-gradient field.{\large \ }

It was shown in Ref. \cite{L-field} that the main particle production occurs
in an inner subrange $\tilde{\Omega}_{K}$\ of the Klein zone, $\tilde{\Omega}%
_{K}\subset \Omega _{K}$,{\large \ }%
\begin{align}
& \tilde{\Omega}_{K}:\ |p_{0}|/\sqrt{eE}<\sqrt{eE}L/2-K,\ \lambda <K_{\bot
}^{2},  \notag \\
& \lambda =\frac{\mathbf{p}_{\bot }^{2}+m^{2}}{eE},\ \sqrt{eE}L\gg K\gg
K_{\bot }^{2}\gg \max \{1,E_{c}/E\}.  \label{b}
\end{align}%
where $K$ and $K_{\bot }$\ are any given positive numbers satisfying the
condition (\ref{b}).

The differential number of particles with quantum numbers $n\in $ $\tilde{%
\Omega}_{K}$ created from the vacuum reads 
\begin{align}
& N_{n}^{\text{\textrm{cr}}}=e^{-\pi \lambda }\left[ 1+O(|\xi
_{1}|^{-3})+O\left( |\xi _{2}|^{-3}\right) \right] ,  \notag \\
& \xi _{1}=\frac{-eEL/2-p_{0}}{\sqrt{eE}},\ \xi _{2}=\frac{eEL/2-p_{0}}{%
\sqrt{eE}}.  \label{6.4}
\end{align}%
We recall that, in fact,\emph{\ }the quantum numbers $n$ that label electron
and positron states in general formulas gather several quantum numbers,%
\begin{equation}
n=\left( p_{0},\mathbf{p}_{\perp },\sigma \right) ,\text{ \ }\mathbf{p}%
_{\perp }=\left( p_{2},\ldots ,p_{D}\right) ,  \label{6.3}
\end{equation}%
where for an electron $p_{0}$\ is its energy and for a positron $-p_{0}$\ is
its energy, for an electron\emph{\ }$\mathbf{p}_{\perp }$\emph{\ }denote its
transversal components of the momentum, whereas for\emph{\ }a positron\emph{%
\ }$-\mathbf{p}_{\perp }$\ denote its transversal components of the
momentum. For an electron $\sigma $\ is its spin polarization and for a
positron $-\sigma $\ is its\emph{\ }spin polarization. Note that the
electron and positron in a pair created by an external field have the same
quantum numbers $n$.

The quantity (\ref{6.4}) is almost constant over the wide range of energy $%
p_{0}$ for any given $\lambda $ $<K_{\bot }^{2}$, for these quantum numbers
we can assume $N_{n}^{\text{\textrm{cr}}}\approx e^{-\pi \lambda }$. In the
limiting case of the large work approximation, $\sqrt{eE}L\rightarrow \infty 
$, one obtains the well-known result for particle creation by a constant
uniform electric field $N_{n}^{\text{\textrm{cr}}}=e^{-\pi \lambda }$, see
Ref. \cite{Nikishov1,Nikishov2,Nikishov3}.

In the approximation under the consideration, the total number of  particles
created from the vacuum is given by a sum (integral) over $n\in \tilde{\Omega%
}_{K}$,%
\begin{equation}
N^{\text{$\mathrm{cr}$}}=\sum_{n\in \Omega _{K}}N_{n}^{\text{$\mathrm{cr}$}%
}\approx \sum_{\mathbf{p}_{\bot },\text{ }p_{0}\in \tilde{\Omega}%
_{K}}\sum_{\sigma }N_{n}^{\text{$\mathrm{cr}$}}=\frac{J_{(d)}TV_{\bot }}{%
(2\pi )^{d-1}}\int_{\tilde{\Omega}_{K}}dp_{0}d\mathbf{p}_{\bot }N_{n}^{\text{%
$\mathrm{cr}$}}\ .  \label{6.6}
\end{equation}%
where $J_{(d)}=2^{\left[ d/2\right] -1}$ is a spin summation factor, $%
V_{\bot }$ is the $(d-2)$-dimensional spatial volume in hypersurface
orthogonal to the electric field direction and $T$ is the time duration of
the electric field. The integration over $p_{0}$ results in 
\begin{equation}
N^{\text{$\mathrm{cr}$}}=\frac{J_{(d)}TV_{\bot }LeE}{(2\pi )^{d-1}}\int_{%
\tilde{\Omega}_{K}}d\mathbf{p}_{\bot }e^{-\pi \lambda }\text{ }.  \label{6.8}
\end{equation}%
Integrating Eq. (\ref{6.8}) over $p_{\bot }$, we obtain that the total
number of created from the vacuum particles in the large work approximation
has the form%
\begin{equation}
N^{\text{$\mathrm{cr}$}}=\frac{J_{(d)}TV(eE)^{d/2}}{(2\pi )^{d-1}}\exp
\left( -\pi \frac{E_{c}}{E}\right) ,  \label{6.9}
\end{equation}%
where $V=$ $LV_{\bot }$ is the volume inside of the capacitor (the volume
occupied by the electric field).

It is obvious that $N^{\text{\textrm{cr}}}<\infty $, when the values $V$ and 
$T$ are finite, or, in other words, when regularization of the finite volume
and finite time of the field action is used. Looking on the condition (\ref%
{2.12}), we see that the $x$-electric potential step which represent the
electric field inside of the capacitor does not violate the unitarity in QED.

Let us estimate the information loss of the reduced states of the deformed
vacuum, which can be calculated as entropies (\ref{4.13}) of these states,.
Using the same summation rule as in (\ref{6.6}), one can write 
\begin{equation}
S(\hat{\rho}_{\pm })=-k_{B}\frac{J_{(d)}TV_{\bot }}{(2\pi )^{d-1}}%
\int_{\Omega _{K}}dp_{0}d\mathbf{p}_{\bot }\left[ N_{n}^{\text{$\mathrm{cr}$}%
}\ln N_{n}^{\text{$\mathrm{cr}$}}+(1-N_{n}^{\text{$\mathrm{cr}$}})\ln
(1-N_{n}^{\text{$\mathrm{cr}$}})\right] .  \label{6.10}
\end{equation}

For Fermi particles under the consideration, $N_{n}^{\text{\textrm{cr}}}\leq
1$. This allows us to expand the logarithm in the RHS of Eq. (\ref{6.10}) in
powers of $N_{n}^{\text{\textrm{cr}}}$. Thus, we represent the term $%
(1-N_{n}^{\text{\textrm{cr}}})\ln (1-N_{n}^{\text{\textrm{cr}}})$ as follows 
\begin{equation}
(1-N_{n}^{\text{$\mathrm{cr}$}})\ln (1-N_{n}^{\text{$\mathrm{cr}$}})=-\
(1-N_{n}^{\text{$\mathrm{cr}$}})\sum_{l=1}^{\infty }l^{-1}\left( N_{n}^{%
\text{$\mathrm{cr}$}}\right) ^{l}.  \label{6.11}
\end{equation}%
Using (\ref{6.11}) in Eq. (\ref{6.10}), we obtain the following intermediate
result 
\begin{equation}
S(\hat{\rho}_{\pm })=k_{B}\frac{J_{(d)}TV_{\bot }}{(2\pi )^{d-1}}%
\int_{\Omega _{K}}dp_{0}d\mathbf{p}_{\bot }\left[ -N_{n}^{\text{$\mathrm{cr}$%
}}\ln N_{n}^{\text{$\mathrm{cr}$}}+(1-N_{n}^{\text{$\mathrm{cr}$}%
})\sum_{l=1}^{\infty }l^{-1}\left( N_{n}^{\text{$\mathrm{cr}$}}\right) ^{l}%
\right] .  \label{6.12}
\end{equation}

As we have mentioned before, the considerable amount of particles is created
only in the subrange $\tilde{\Omega}_{K}\in \Omega _{K}$, where terms
proportional to $|\xi _{1,2}|^{-3}$ are small and can be neglected, allowing
to use the leading-order approximation $N_{n}^{\text{\textrm{cr}}}\approx
e^{-\pi \lambda }$ in the RHS of Eq. (\ref{6.12}). Then we obtain 
\begin{eqnarray}
S(\hat{\rho}_{\pm }) &\approx &k_{B}\frac{J_{(d)}TVeE}{\left( 2\pi \right)
^{d-1}}\int_{\tilde{\Omega}_{K}}d\mathbf{p}_{\bot }\left[ \pi \lambda
e^{-\pi \lambda }+(1-e^{-\pi \lambda })\sum_{l=1}^{\infty }l^{-1}e^{-\pi
\lambda l}\right] \text{ }\;\mathrm{if}\;d>2;  \notag \\
S(\hat{\rho}_{\pm }) &\approx &k_{B}\frac{TVeE}{2\pi }A\left(
2,E_{c}/E\right) \text{ }\;\mathrm{if}\;d=2,  \notag \\
A\left( 2,E_{c}/E\right)  &=&\left\{ \pi E_{c}/E\exp \left( -\pi
E_{c}/E\right) -\left[ 1-\exp \left( -\pi E_{c}/E\right) \right] \ln \left[
1-\exp \left( -\pi E_{c}/E\right) \right] \right\} .  \label{6.13}
\end{eqnarray}

In the dimensions $d>2$ the integration over the transversal components of
the momentum can be easily performed. Outside of the subrange $\tilde{\Omega}%
_{K}$, the integrand is very small, so that we can extend the integration
limits of $p_{\bot }$\ to the infinity. Thus, we finally get 
\begin{equation}
S(\hat{\rho}_{\pm })\approx k_{B}\frac{J_{(d)}TV(eE)^{d/2}}{(2\pi )^{d-1}}%
A\left( d,E_{c}/E\right) \text{ }\;\mathrm{if}\;d>2,  \label{6.14}
\end{equation}%
where the factor $A\left( d,E_{c}/E\right) $ has the form%
\begin{eqnarray}
&&\ A\left( d,E_{c}/E\right) =\left( \pi E_{c}/E+d/2-1\right) \exp \left(
-\pi E_{c}/E\right)   \notag \\
&&+\sum_{l=1}^{\infty }\left[ l^{-d/2}-l^{-1}(l+1)^{(2-d)/2}\exp \left( -\pi
E_{c}/E\right) \right] \exp \left( -\pi lE_{c}/E\right) .  \label{6.15}
\end{eqnarray}%
For example, estimations of this factor for strong field $E_{c}/E\ll 1$\ and
critical field $E_{c}/E=1$\ with $d=4,3$\ are $A\left( 4,0\right) =\pi ^{2}/6
$, $A\left( 4,1\right) \approx 0,22$; $A\left( 3,0\right) \approx 0,93$, $%
A\left( 3,1\right) \approx 0,20$. In the case of a weak field,  $E_{c}/E\gg 1
$, the entropy is exponentially small for any $d$,%
\begin{equation*}
A\left( d,E_{c}/E\right) \approx \left( \pi E_{c}/E+d/2\right) \exp \left(
-\pi E_{c}/E\right) .
\end{equation*}

One can note, that the large work approximation (\ref{6.14}) obtained for $S(%
\hat{\rho}_{\pm })$ in the case of the $x$-electric step under consideration
coincides with the same approximation for $S(\hat{\rho}_{\pm })$ in the case
of the $t$-electric step with an uniform electric field that is acting
during a finite time interval $T$ (the so called $T$-constant field)
obtained in Ref. \cite{GavGitSh15}. This observation confirms the fact that
the $T$-constant and $L$-constant fields produce equal physical effects in
the large work approximation (or as $T\rightarrow \infty $ and $L\rightarrow
\infty $), such that it is possible to consider these fields as
regularizations of a constant uniform electric field given by two distinct
gauge conditions for electromagnetic potentials. Obviously, exact
expressions for the entropies $S(\hat{\rho}_{\pm })$ differ in the general
case.

\section*{Acknowledgments}

The work of the authors{\large \ }was supported by a grant from the Russian
Science Foundation, Research Project No. 15-12-10009.

\end{document}